\documentclass[preprintnumbers,showpacs,prl,twocolumn]{revtex4}
\usepackage[dvips]{texdraw}
\usepackage{amssymb}
\usepackage{mathrsfs}

\newcommand{\AdSxS}{AdS$_5\times{\mathcal S}^5$}

\begin{document}
\preprint{IFUM-812-FT}
\title{Chronology Protection in anti-de~Sitter}
\author {Marco M.~Caldarelli}
\email{marco.caldarelli@mi.infn.it}
\author{Dietmar Klemm}
\email{dietmar.klemm@mi.infn.it}
\author{Pedro J. Silva}
\email{pedro.silva@mi.infn.it}
\affiliation{Dipartimento di Fisica dell'Universit\`a di Milano}
\affiliation{INFN, Sezione di Milano,\\
Via Celoria 16, I-20133 Milano.}
\begin{abstract}
We consider 1/2 BPS excitations of AdS$_5$ $\times$ ${\cal S}^5$
geometries in type IIB string theory that can be mapped into free fermion
configurations according to the prescription of Lin, Lunin and
Maldacena (LLM). It is shown that whenever the fermionic probability density
exceeds one or is negative, closed timelike curves appear in the bulk.
A violation of the Pauli exclusion principle in the phase space of the fermions is thus intimately related to causality violation in the dual geometries.
\end{abstract}
\maketitle


Closed timelike curves (CTCs) plague solutions of general relativity and supergravity theories. In contrast to curvature singularities, which may be coped with by introducing particular boundary conditions or by invoking some more fundamental theory modifying the behaviour of the gravitational force near the singularity, CTCs are more subtle. They are not linked to the presence of strong gravitational fields, and
the resulting spacetime is not globally hyperbolic.
Moreover, they are global in nature, making it difficult to recognise them and to capture their presence from the local field equations of gravity. This problem was first raised by Kurt G\"odel \cite{Godel:1949ga}, who found a homogeneous and simply connected universe with
CTCs through every point of spacetime. This particular solution is not an isolated pathology and in fact, 
it was soon recognized that CTCs 
appear frequently in
gravitational theories, in particular in higher dimensions (see for example \cite{Myers:1986un}). Supersymmetry does not help to rule out chronological pathologies \cite{Gibbons:1999uv}, nor does the lifting to higher dimensional supergravities \cite{Caldarelli:2001iq}.

The usual way out is to simply label spacetimes with CTCs as unphysical, and to forget about them.
However, time machines can be constructed from initial data on spatial hypersurfaces whose causal past does not contain CTCs \cite{Gott:1990zr,Cutler:1992nn}.
This lead Hawking to formulate the chronology protection conjecture \cite{Hawking:1991nk}, with a quantum mechanism enforcing it by superselecting the causality violating field configurations from the quantum mechanical phase space \cite{Cassidy:1997kv}.

Here, we shall take a more fundamental point of view, and consider the (super-)gravity theories merely as effective low energy limits of string theory. Then, even if the effective theory cannot distinguish between causally well-behaved and ill spacetimes, one can hope that string theory provides a mechanism to rule out the pathological solutions or eliminate the CTCs. This approach has been successful in the case of the BMPV black hole \cite{Breckenridge:1996is}, which develops CTCs in the overrotating regime. Using the microscopic description in terms of a two-dimensional conformal field theory, provided by string theory, Herdeiro showed that the causality bound corresponds to a unitarity bound in the CFT \cite{Herdeiro:2000ap}, and therefore that the overrotating spacetimes are not genuine solutions of string theory. A similar mechanism was proposed in \cite{Caldarelli:2001iq} to hold for \AdSxS{} solutions of type IIB supergravity. In this case, using the AdS/CFT correspondence \cite{Maldacena:1997re}, we have a non-perturbative formulation of type IIB superstring theory in terms of ${\mathcal N}=4$ SYM in four dimensions.
In particular, thanks to the work by Lin, Lunin and Maldacena \cite{Lin:2004nb}, 
we have a strong control on a subsector of half BPS configurations.
In this subsector, the SYM theory reduces to a system of $N$ free fermions in a harmonic well, whose classical configurations are given by quantum Hall droplets in their phase space, i.e.~by the description of a region of area $\hbar N$ where the fermions are localized \footnote{The Fermi cell has area $\hbar=2\pi l^4_p$, with $l_p$ the Planck length, and the radius $L$ of AdS$_5$ is related to them by $L^4=4\pi l^4_pN$.}. LLM define a function $z(x^1,x^2,0)$ which takes the value $-1/2$ inside the Hall droplets, and $1/2$ outside. The gravitational dual to this configuration is then given by the metric \footnote{For the five-form field strength we refer to the original paper \cite{Lin:2004nb}.}
\begin{eqnarray}
	ds^2=-h^{-2}\left(dt+V_idx^i\right)^2+h^2\left(dy^2+dx^idx^i\right)
	\nonumber\\
		+ye^Gd\Omega_3^2+ye^{-G}d\tilde\Omega_3^2\,,
\label{LLM}\end{eqnarray}
with
\begin{equation}
	h^{-2}=\frac{2y}{\sqrt{1-4z^2}}\,,\qquad e^{2G}=\frac{1+2z}{1-2z}\,.
\end{equation}
The function $z(x^i,y)$ and the shift vector $V_i$ are determined by fermion configurations $z(x^i,0)$, which act as boundary value,
\begin{equation}
	z(x^1, x^2, y)=\frac{y^2}\pi\int\frac{z(x'_1,x'_2,0)\,dx'_1\,dx'_2}{\left[
	\left({\bf x}-{\bf x}'\right)^2+y^2\right]^2}\,,
\end{equation}
\begin{equation}
V_i(x^1,x^2,y)=\frac{\epsilon_{ij}}\pi\int\frac{z(x'_1,x'_2,0)(x_j-x_j')\,dx'_1\,dx'_2}{\left[\left({\bf x}-{\bf x}'\right)^2+y^2\right]^2}\,.
\end{equation}
We have therefore a very nice realization of holography, in which the dual field theory data is mapped on the $y=0$ two-plane of the gravity solution.

It is useful to describe the Fermi liquid using the variable $\rho(x^1,x^2)=1/2-z(x^1,x^2,0)$, which takes the value one inside the hall droplets, and zero outside. Then, it is tempting to go beyond the classical limit, and interpret the function $\rho$ as the probability density of the fermions. From the point of view of the SYM theory, as long as $\rho\in[0,1]$, one has a well-defined semi-classical state of the field theory yielding this probability density. 
Due to the noncommutativity of the phase space variables, the variation of $\rho
$ must be slow on the scale of $\sqrt\hbar$.
A $\rho>1$ would correspond to a violation of Pauli's exclusion principle, while negative density clearly has no meaning. However, on the gravitional side of the correspondence, {\em any} real function $\rho$ gives a one half BPS solution of the supergravity equations of motion, in general with some singularities. In this letter, we shall show that 
whenever the above physical conditions on $\rho$ are violated,
the dual spacetime has closed timelike curves. Therefore, we propose that {\it the requirement of unitarity of the dual CFT implements the chronology protection for superstring theory on
asymptotically \AdSxS spacetimes}.
We shall first show this behaviour in the case of the superstar solution, and then give a general proof for rotationally symmetric Hall droplets.

\section{LLM description of the superstar}

The simplest family of half BPS solutions of type IIB supergravity on \AdSxS{} is given by the so-called superstar solution
\begin{eqnarray}
ds^2=-\frac1{\sqrt{\Delta}}\left(\cos^2\theta+\frac{r^2}{L^2}\Delta\right)dt^2+
\frac{L^2H}{\sqrt\Delta}\sin^2\theta\,d\phi^2\nonumber\\
+\frac{2L}{\sqrt\Delta}\sin^2\theta\,dt\,d\phi
+\sqrt\Delta\left(\frac{dr^2}{f}+r^2\,d\Omega_3^2\right)\nonumber\\
+L^2\sqrt\Delta\,d\theta^2+\frac{L^2}{\sqrt\Delta}\cos^2\theta\,d\tilde\Omega_3^2\,,\qquad\qquad
\label{superstar}\end{eqnarray}
with $H=1+Q/r^2$, $f=1+Hr^2/L^2$ and $\Delta=\sin^2\theta+H\cos^2\theta$. The ${\mathcal S}^5$ factor of the metric in these coordinates is given by
$d\theta^2+\sin^2\theta\,d\phi^2+\cos^2\theta\,d\tilde\Omega_3^2$.
This solution is obtained by oxidation to ten dimensions \cite{Cvetic:1999xp} of the extremal limit of the black hole solution with one R-charge of the $STU$ model \cite{Behrndt:1998ns,Behrndt:1998jd}. Again, there is a five-form field strength turned on, which can be found in \cite{Cvetic:1999xp}. For $Q=0$ we recover standard \AdSxS.
The $Q>0$ case exhibits a naked singularity at the origin $r=0$ of AdS, where a condensate of giant gravitons growing in the five-sphere sits, and acts as a source for the supergravity fields \cite{Myers:2001aq}. This view has been confirmed in \cite{Lin:2004nb}, where the corresponding fermion distribution in the dual CFT was interpreted as a dilute gas of holes in the Fermi sea. In fact, using the coordinate transformation
\begin{equation}
y=Lr\cos\theta\,,\qquad R=L^2\sqrt{f}\sin\theta\,,\qquad t\mapsto Lt\,,
\end{equation}
one recovers the LLM metric (\ref{LLM}) with polar coordinates $(R,\phi)$ on the $(x^1,x^2)$ plane. With some algebra, one can then compute the corresponding function $z(R,\phi,y)$, that reads
\[
z={1\over 2(1+Q/L^2)}\left[ {y^2+R^2-R_0^2\over \sqrt{(y^2+R^2+R_0^2)^2-4R^2R_0^2}}+{Q\over L^2}\right]\,,
\]
where $R_0^2=L^2(L^2+Q)$. Its $y=0$ limit yields the fermion distribution
\begin{equation}
	\rho(R) = \left\{\begin{array}{c@{\qquad}l}
		\displaystyle\frac1{1+Q/L^2} & R < R_0\\
			                         &	                \\
		0                            & R > R_0\,.
	\end{array}\right.
\label{superstar-density}\end{equation}
LLM showed that the vacuum \AdSxS{} is represented by a Fermi droplet of density $\rho=1$ and radius $L^2$. Its total area $\pi L^4$ consists therefore exactly of $N$ Fermi cells of area $\hbar=2\pi l_p^4$. By turning on the R-charge $Q$, the probability density spreads to a disk of radius
$L^2\sqrt{1+Q/L^2}$,
but with lower density, in such a way that the correct number of fermions $\frac{1}\hbar\int\!\rho\,da=N$ is recovered. This fact supports the interpretation of $\rho$ as a density distribution of fermions.
The fermion system represents a uniform gas of holes delocalized in the Fermi sea, and since such holes correspond to giant gravitons growing on the five-sphere
\cite{Berenstein:2004kk,Caldarelli:2004ig}, the superstar can be thought of as the backreaction on spacetime produced by a condensate of giant gravitons \cite{Myers:2001aq}.

\begin{figure}
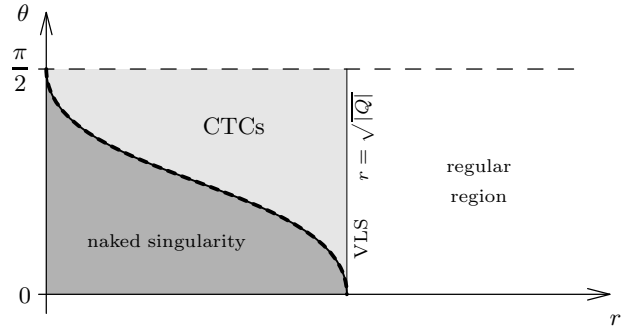

        \begin{center}
        \begin{texdraw}
        \drawdim cm \setunitscale 1
\move (-4 -1.5)\lvec(-4 1.5)\lvec(0 1.5)\lvec(0 -1.5)\ifill f:.9
\move(-4 1.5)
\clvec(-4 0)(0 0)(0 -1.5)
\lvec(-4 -1.5)\ifill f:.7
\linewd 0.025
\move (-4 1.5)
\clvec(-4 0)(0 0)(0 -1.5)
\linewd 0.05
\lpatt(0.1 0.1)
\move (-4 1.5)
\clvec(-4 0)(0 0)(0 -1.5)
\lpatt()
\htext{\scriptsize}
\linewd 0.02
\arrowheadsize l:0.3 w:0.15 \arrowheadtype t:V
\move (-4.1 -1.5) \avec (3.5 -1.5)
\htext(3.5 -1.9){$r$}
\move (-4 -1.75) \avec (-4 2.25)
\textref h:C v:C \htext(-4.3 2.25){$\displaystyle\theta$}
\linewd .02 \lpatt(.2 .2)
\textref h:R v:C \htext(-4.2 -1.5){$\displaystyle 0$}
\move (-4.1 1.5) \lvec (3.2 1.5)
\textref h:R v:C \htext(-4.2 1.5){$\displaystyle\frac\pi2$}
\linewd .01 \lpatt()
\move(0 -1.5)\lvec(0 1.5)
\textref h:C v:T \vtext(.05 .6){\scriptsize $r=\sqrt{|Q|}$}
\vtext(.1 -.75){\scriptsize VLS}
\textref h:C v:C \htext(-1.5 .75){CTCs}
\textref h:C v:C \htext(-2.4 -.8){\scriptsize naked singularity}
\textref h:C v:C \htext(1.75 .2){\scriptsize regular}
\textref h:C v:C \htext(1.75 -.2){\scriptsize region}
\end{texdraw}
\end{center}
	\caption{Location of the naked singularity, the region with CTCs and the velocity of light surface (VLS) for the anti-superstar with $-L^2<Q<0$.}
\label{fig-superstar}\end{figure}

Let us suppose now that $-L^2<Q<0$ \footnote{The case with $Q<-L^2$ corresponds to putting sources away from the $(x^1, x^2)$-plane, and is therefore not covered by the LLM construction.}. The metric (\ref{superstar}) still represents a genuine solution of the supergravity theory. This spacetime is defined for
$r>\sqrt{|Q|}\cos\theta$, and has a naked singularity at $r=\sqrt{|Q|}\cos\theta$.
For $r>\sqrt{|Q|}$ it is perfectly well-behaved, but one checks that in-between,
for $\sqrt{|Q|}\cos\theta<r<\sqrt{|Q|}$, the coefficient $g_{\phi\phi}$ of the metric becomes negative, and the orbits of $\partial_\phi$ are CTCs (see figure~\ref{fig-superstar}).
Note that the metric remains lorentzian everywhere.
As in the superstar case, we can cast the metric in the LLM form to read off the source which produces this spacetime. The result is again (\ref{superstar-density}), but this time, since $Q<0$, the Fermi sea has been squeezed to a smaller radius, resulting in a density $\rho>1$. This implies the presence of more than one fermion per Fermi cell, in plain violation of Pauli's exclusion principle. The anti-superstar has no dual CFT state. Taking the dual CFT description equivalent to type IIB superstring theory as the fundamental one,
we see that the Pauli exclusion principle automatically rules out the solutions of the form (\ref{superstar}) with causal anomalies.
In the following we shall argue that this mechanism works in much more generality.

\section{Chronology protection and Pauli exclusion principle}

We shall, for simplicity, restrict to circular symmetric configurations of fermions.
In this case, $\rho=\rho(R)$
(where $\rho(x^i)=1/2-z(x^i,0)$) and
\begin{equation}
z(R,y)=2y^2\int_0^\infty\frac{z(r,0)\left(R^2+y^2+r^2\right)r\,dr}{\left[\left(R^2+r^2+y^2\right)^2-4R^2r^2\right]^{3/2}}
\end{equation}
is a continuous function for $y>0$.
The integral converges if $z(R,0)$ grows slower than $R^2$ at infinity.
However, to have a finite number $N=\frac1\hbar\int\!\rho\,da$ of fermions, the density $\rho(R)$ must decay faster than $1/R^2$ at infinity \footnote{By relaxing this assumption one would obtain more general asymptotic behaviors.}. Therefore,
\begin{equation}
	z_\infty\equiv\lim_{R\rightarrow\infty}z(R,0)=\frac12
\end{equation}
and the supergravity solution is asymptotically \AdSxS. Using this assumption, one obtains
\begin{equation}
\lim_{y\rightarrow\infty}z(R,y)=z_{\infty}\,.
\end{equation}
It is also easy to prove that if $|z(R,0)|\leq1/2$ (corresponding to a good fermionic configuration since $\rho\in[0,1]$), then $|z(R,y)|<1/2$ everywhere, while points $(R,y)$ for which $|z(R,y)|>1/2$ do not belong to the spacetime;
there the metric becomes imaginary and to reach them one has to cross a curvature singularity located in $z=\pm1/2$ \footnote{To check the occurrence of the curvature singularity one can approximate the metric near the hypersurface $z=\pm1/2$ keeping just the leading terms in $Y=|z|-1/2$. We thank M.~O'Loughlin for pointing this out.}.

Next, we observe that if the function $z$ takes the value $\pm1/2$ at some point $(R_0,y_*)$ with $y_*\neq0$, then there are CTCs in its neighborhood (if there are points with $|z|<1/2$ in its neighborhood). 
Indeed, since
\begin{equation}
g_{\phi\phi}=\frac{-2y}{\sqrt{1-4z^2}}\left(V_\phi^2-\frac{R^2}{4y^2}(1-4z^2)\right)\,,
\end{equation}
one can check that keeping $R_0$ fixed and supposing $|z(R_0,y)|<1/2$ for $y>y_*$, $\lim_{y\to y_*^+}g_{\phi\phi}=-\infty$
\footnote{Strictly speaking, this holds only if $V_\phi\neq0$. However, one can check that the directional derivative of $V_\phi$ along the curves $z=\pm1/2$ in the $(R,\phi)$ plane is not everywhere vanishing, and therefore if $V_\phi(R,y_*(R))=0$, one can always choose a point $(R+\epsilon,y_*(R+\epsilon))$ where $V_\phi\neq0$.}.
Therefore, there exist a $y_0>y_*(R_0)$ such that $g_{\phi\phi}<0$, and the orbits of $\partial_\phi$ through this point $(R_0,y_0)$ are CTCs.

Let us suppose that in some points of the phase space the fermion density violates the exclusion principle, that is $\rho(R)>1$ for some values of the radius, or equivalently $z(R,0)<-1/2$. There, the spacetime is not defined, because as $y$ decreases
one encounters a singularity at some point.
Let us keep $R$ constant; the function $z(R,y)$ is continuous in $y$ and takes the value $z_\infty>0$ for $y\rightarrow\infty$. Therefore, if $z(R,0)<-1/2$, there is some point $y_*$ for which $z(R,y_*)=-1/2$. Let us define the function $y_*(R)$,
\[
	y_*(R)=\left\{\begin{array}{c@{\qquad}l}
		\max\left\{y|z(R,y)=-\frac12\right\}\,, & z(R,0) < -\frac12\\
		\\
		0\,, & |z(R,0)| \leq \frac12\\
		\end{array}\right.\nonumber
\]
Then, the region $y>y_*(R)$ of the spacetime defines a good supergravity solution, with a singularity in $y=y_*(R)$, and by the previous observation it follows that one can construct CTCs in the neighborhood of the singularity. A similar argument holds if we have a negative probability density in some region of the phase space. Since the asymptotic value of $z$ for large $y$ is $1/2$, there must be by continuity of the function $z$ some point with nonvanishing $y$ such that $z=1/2$, yielding CTCs in its neighborhood.

Finally, since $|z(R,y)|<1/2$ for $y>0$ if $|z(R,0)|\leq1/2$, it is unlikely that CTCs arise when $\rho$ gives a good probability density, but unfortunately we were unable to prove this.

\section{Discussion}

We have shown that if the boundary condition $\rho(R)$ for the supergravity solution violates somewhere the bound $0\leq\rho(R)\leq1$, the spacetime suffers from CTCs. If we interpret $\rho$ as the probability density of fermions in the phase space, this bound corresponds to a unitarity bound of the dual field theory; $\rho>1$ would violate Pauli's exclusion principle, while negative $\rho$ would imply negative norm states. Therefore, such configurations do not correspond to any state in the Hilbert space of the SYM theory, which gives a fundamental description of superstring theory on \AdSxS, and the corresponding causality-violating backgrounds are forbidden. This mechanism implements in a simple form 
the chronology protection conjecture for a subsector of type IIB superstring theory in asymptotic \AdSxS{} backgrounds, and is similar to the one acting for the BMPV black hole \cite{Herdeiro:2000ap}. 
Stronger evidence would come by proving that no CTCs arise for boundary conditions verifying $0\leq\rho(R)\leq1$; although we didn't find any counterexample, we were not able to prove this in general.
The coherence of the whole picture provides some further support to the interpretation of $\rho(R)$ as a probability density. 

This interpretation of $\rho$ allows then the study of the half BPS sector of AdS/CFT beyond the strictly classical supergravity limit, and the study of thermal states.
Using the fact that right and left movers of the matrix model do not interact, one can thermally excite the left-movers, giving a temperature to the fermion gas, maintaining the right movers on the ground state and remaining therefore in the 1/2 BPS sector. The corresponding supergravity background has been constructed in \cite{Buchel:2004mc}. It would be interesting to elucidate the link between this state and the stretched horizon which develops for the superstar \cite{Suryanarayana:2004ig}. Finally, it would be interesting to understand how this chronology protection mechanism works in other subsectors of AdS/CFT.


\begin{acknowledgments}
The authors are grateful to S.~L.~Cacciatori, W.~Mueck and M.~O'Loughlin for useful discussions.
This work was partially supported by INFN, MURST and
by the European Commission program MRTN-CT-2004-005104.
\end{acknowledgments}


\end{document}